\documentclass[twocolumn]{aastex63}

\newcommand{\ANADDR}[1]{}

\DeclareMathSymbol{\varOmega}{\mathord}{letters}{"0A}
\DeclareMathSymbol{\varSigma}{\mathord}{letters}{"06}
\DeclareMathSymbol{\varPsi}{\mathord}{letters}{"09}
\DeclareMathSymbol{\varPhi}{\mathord}{letters}{"08}
\DeclareMathSymbol{\varGamma}{\mathord}{letters}{"00}

\received{}
\revised{}
\accepted{}
\submitjournal{ApJ}

\shorttitle{Transport, destruction and growth of pebbles in a gas envelope}
\shortauthors{Johansen \& Nordlund}

\begin{document}

\title{Transport, destruction and growth of pebbles in the gas envelope of a
protoplanet}

\correspondingauthor{Anders Johansen}
\email{anders@astro.lu.se}

\author{Anders Johansen}
\affiliation{Lund Observatory, Lund University\\
Box 43, 22100 Lund, Sweden}
\affiliation{Centre for Star and Planet Formation, Globe Institute, University of Copenhagen \\ \O ster Voldgade 5–7, 1350 Copenhagen, Denmark}

\author{\AA ke Nordlund}
\affiliation{Niels Bohr Institute, University of Copenhagen \\
Blegdamsvej 17, 2100 Copenhagen, Denmark}

\begin{abstract}

We analyse the size evolution of pebbles accreted into the gaseous envelope of a
protoplanet growing in a protoplanetary disc, taking into account collisions
driven by the relative sedimentation speed as well as the convective gas motion.
Using a simple estimate of the convective gas speed based on the pebble
accretion luminosity, we find that the speed of the convective gas is higher
than the sedimentation speed for all particles smaller than 1 mm. This implies
that both pebbles and pebble fragments are strongly affected by the convective
gas motion and will be transported by large-scale convection cells both towards
and away from the protoplanet's surface. We present a simple scheme for evolving
the characteristic size of the pebbles, taking into account the effects of
erosion, mass transfer and fragmentation. Including the downwards motion of
convective cells for the transport of pebbles with an initial radius of 1
millimeter, we find  pebble sizes between 100 microns and 1 millimeter near the
surface of the protoplanet. These sizes are generally amenable to accretion at
the base of the convection flow. Small protoplanets far from the star ($>30$ AU)
nevertheless erode their pebbles to sizes below 10 microns; future
hydrodynamical simulations will be needed to determine whether such small
fragments can detach from the convection flow and become accreted by the
protoplanet.

\end{abstract}


\section{Introduction}

The rapid accretion of millimeter-sized pebbles appears to be a necessary
ingredient in forming the cores of cold gas giants and ice giants within the
life-time of the gaseous protoplanetary disc
\citep{JohansenLambrechts2017,Ormel2017,Johansen+etal2019,JohansenBitsch2019}.
The formation of super-Earths in the inner regions of the protoplanetary disc
may also be driven by accretion from the drifting pebble population
\citep{Lambrechts+etal2019,Izidoro+etal2019}. The fate of the pebbles after
entering the gaseous envelope of a protoplanet is nevertheless relatively poorly
explored and poorly understood. A recent paper concluded that pebbles will be
sandblasted to dust in the envelope of a protoplanet and transported back to the
protoplanetary disc with the convective overshoot \citep{Ali-DibThompson2019}, resulting
in a pebble accretion time-scale of at least 3 million years.

The goal of this paper is to perform an independent analysis of the evolution of
the pebble sizes within the gas envelope. We were particularly interested in the
role of convective gas flows for the dynamics of the pebbles and for collisions
between them. Recent work on the hydrodynamics and radiative transfer of
protoplanetary envelopes has been done using an adaptive mesh to resolve the gas
flow down to the surface of the protoplanet
\citep{Popovas+etal2018,Popovas+etal2019}. These simulations demonstrated that
the convective motion of the gas on the one hand had a major influence on the
dynamics of pebbles, while on the other hand -- given the assumptions used -- did 
not have a significant effect on the pebble accretion rates.
Pebbles that are captured into the envelope are sometimes carried closer to the
protoplanet with downwelling cold gas flows, while sometimes instead carried
away in upwelling hot flows. The average accretion rates of
pebbles onto the protoplanet were observed to be relatively unaffected by the
convection. The smallest pebble sizes of 10 microns considered in \cite{Popovas+etal2019} showed indications of a decreased accretion efficiency for some of the simulations, but this 
could simply indicate fluctuations in the rather low accretion efficiency of such small pebbles or that small dust cannot decouple from the convection flow close to the protoplanet surface.
Importantly, the hydrodynamical studies by \cite{Popovas+etal2018} and \cite{Popovas+etal2019} ignored the
evolution of the pebble size during the transport down to the protoplanet's
surface. We therefore focus in this paper on understanding the size evolution of
the accreted pebbles, using 1-D models that either include or exclude the
convective gas motion.

The paper is organized as follows. In Section \ref{s:envelope} we introduce our
gas envelope model and discuss the physics of pebble capture. In Section
\ref{s:noconvection} we present results for the pebble-to-gas ratio and
fragmentation-limited size of the pebbles in a model that ignores the convective
motion of the gas. The following Section \ref{s:convection} includes the
convective motion. In Section \ref{s:coagulation} we evolve the pebble size
using a simple approach that takes into account erosion, fragmentation and
mass transfer, both including and excluding the convective motion. We
conclude on our results in Section \ref{s:conclusions}. Appendix A contains a calculation of the fate of the pebbles for an earlier and later stage of the protoplanetary disc compared to the nominal case presented in the main paper. Appendix B contains an additional numerical experiment where we consider convective models with fixed fragment mass ratios relative to the gas.

\section{Envelope structure and pebble capture}
\label{s:envelope}

We adopt a 1-D model for the gas envelope of the protoplanet in hydrostatic and
energy balance. We run models of the gas envelopes of accreting protoplanets
using a constant gas accretion rate through the protoplanetary disc of
$\dot{M}_{\rm g} = 3 \times 10^{-8}\,M_\odot\,{\rm yr}^{-1}$  and a constant
viscosity of $\alpha=0.01$.

The protoplanetary accretion disc model is only used to set the column density,
$\varSigma_{\rm g}$, of the disk in which the envelope is embedded, through the
relation $\dot{M}_{\rm g} = 3 \pi \alpha c_{\rm s} H \varSigma_{\rm g}$
\citep{Pringle1981}, where $c_{\rm s}$ is the sound speed of the gas and
$H=c_{\rm s}/\varOmega$ is the local vertical scale-height ($\varOmega$ is the
Keplerian frequency). We consider the mass accretion rate rather than an assumed
column density profile in order to facilitate comparisons to the pebble
accretion models of \cite{Johansen+etal2019} who used an evolving alpha-disc. We
use the cold temperature structure $T = 121\,{\rm K}\,(r/{\rm AU})^{-3/7}$ from
\cite{ChiangYoudin2010}. This yields then a gas column density of approximately
$10^4\,{\rm kg\,m^{-2}}$ at 1 AU, similar to the minimum mass solar nebula, but
with a shallow radial logarithmic slope of -15/14. This evolution stage of the
protoplanetary disc is chosen as it constitutes the main growth phase of the
cores of giant planets \citep{Johansen+etal2019}. Planets could form at even
earlier stages of the protoplanetary disc, with accretion rates in the range
between $10^{-6}\,M_\odot\,{\rm yr}^{-1}$ to $10^{-7}\,M_\odot\,{\rm yr}^{-1}$
\citep{Manara+etal2018}. We present the results of considering either an earlier
or a later evolutionary stage of the protoplanetary disc in Appendix A.

We run simulations with planetary masses $M=$ $0.1$, $0.3$, $1.0$, $3.0$ $M_{\rm
E}$ placed at distances $a=$ $1.0$, $3.0$, $10.0$ and $30.0$ AU from the central
star. The pebble accretion time-scale $\tau$ is set to a fixed value of $10^6$
yr. We fix the incoming pebble size to 1 millimeter.

\subsection{Protoplanet envelope}

We calculate the structure of the gas envelope by integrating the gas density
and temperature inwards from the Hill radius down to the planetary surface,
setting the density and temperature at the outer boundary equal to the
protoplanetary disc conditions at the relevant distance. The temperature
gradient is set to be the minimum of the radiative and the convective gradient,
using the opacity power-laws of \cite{BellLin1994} and an ideal gas equation
of state with constant adiabatic index $\gamma=1.4$.  We use here an opacity
corresponding to micron-sized dust with 1\% mass relative to the gas and ignore
for simplicity any increase in mean molecular weight and release of latent heat at the ice sublimation line (at a temperature of 170 K) and the silicate
dust sublimation line (at temperatures of 2,000--3,000 K). We refer to \cite{Chambers2017}, \cite{Brouwers+etal2018} and
\cite{BrouwersOrmel2020} for the effect of the release of water vapour and
silicate vapour on the structure of the envelope.
\begin{figure*}
  \begin{center}
    \includegraphics[width=0.45\linewidth]{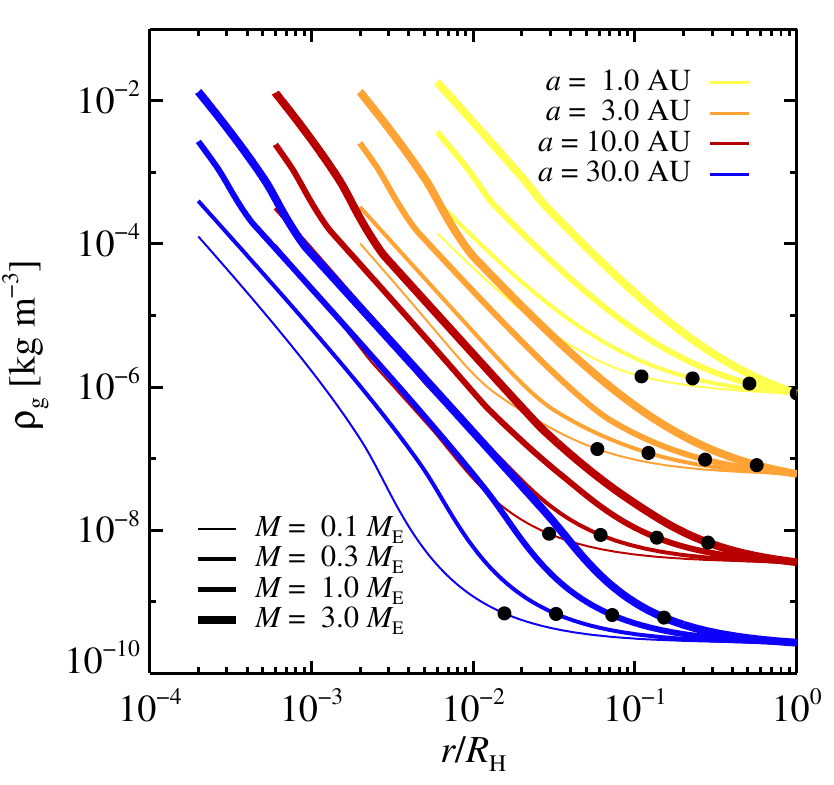}
    \quad
    \includegraphics[width=0.45\linewidth]{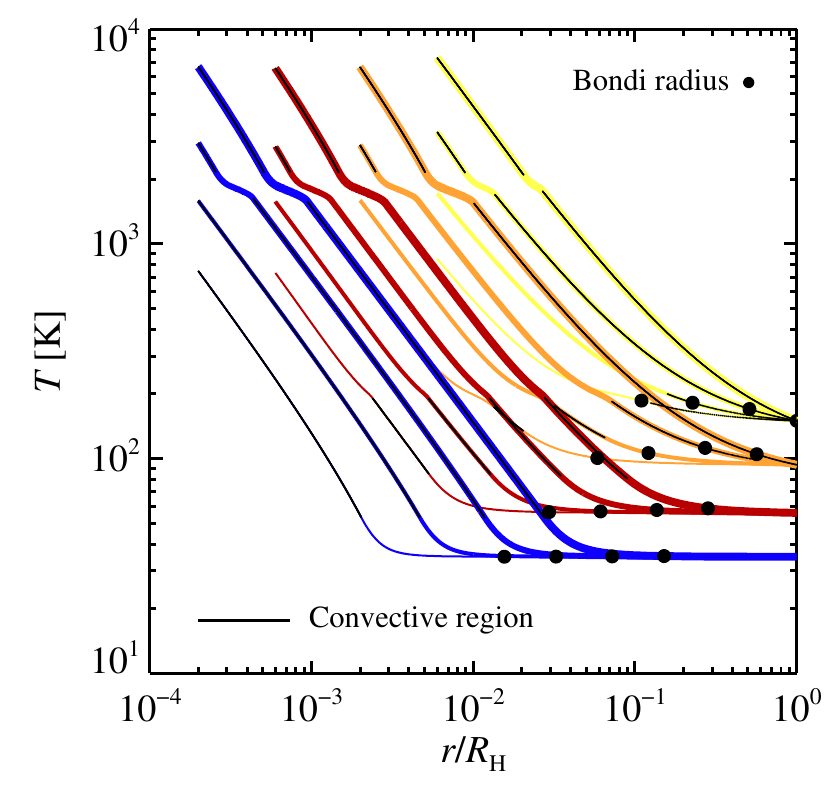}
  \end{center}
  \caption{The gas density (left) and the gas temperature (right) in the
  envelope as a function of the distance from the protoplanet (measured relative
  to the Hill radius). The boundary conditions of the protoplanetary disc in
  terms of density and temperature are matched at $r=R_{\rm H}$. The position of
  the Bondi radius marked, with black dots, relative to the Hill radius
  indicates approximately the protoplanet mass relative to the pebble isolation
  mass. The convective regions are marked with thin black lines in the
  temperature plot. The convectively stable region appearing around 2,000 K is
  due to silicate sublimation in the \cite{BellLin1994} opacity.}
  \label{f:envelope}
\end{figure*}

The resulting envelope structure is shown in Figure \ref{f:envelope}. Both the
density and the temperature display wiggles in their profiles; this is due to
opacity transitions in the \cite{BellLin1994} opacity. Increasing the
protoplanet mass leads to an increase in both the temperature and the density of
the envelope. The temperature nevertheless only crosses the silicate sublimation
temperature near the surfaces of the highest-mass planets. We mark the regions
of convective heat transport in the temperature profile of Figure
\ref{f:envelope}. The higher-mass protoplanets have convective heat transport
through the bulk of the envelope, while some lower-mass protoplanets are not
formally fully convective. Particularly, the sublimation of silicates at
approximately 2,000 K yields a decrease in the opacity with increasing
temperature and hence stability against convection. We will nevertheless
consider the envelopes to be either fully radiative or fully convective in the
analysis of the pebble size evolution, to highlight the differences between
these two extreme cases.

\subsection{Pebble capture}

Pebbles that pass the protoplanet with the Keplerian shear flow are captured
with the help of gas drag when their terminal velocity is approximately equal to
the Keplerian shear speed \citep{OrmelKlahr2010,LambrechtsJohansen2012},
\begin{equation}
  \tau_{\rm f} \frac{G M}{b^2} \sim \varOmega b \, .
\end{equation}
Here $\tau_{\rm f}$ is the friction time of the pebbles and $b$ is the impact
parameter along the radial axis from the central star. The maximum impact
parameter for accretion is called the accretion radius $R_{\rm acc}$. A more
precise analysis \citep{Morbidelli+etal2015} yields the pebble accretion
radius as a function of the Stokes number ${\rm St} = \varOmega \tau_{\rm f}$
and Hill radius $R_{\rm H} = [G M/(3 \varOmega^2)]^{1/3}$ as
\begin{equation}
  R_{\rm acc} = \left( \frac{{\rm St}}{0.1} \right)^{1/3} R_{\rm
  H} \, .
  \label{eq:Racc}
\end{equation}
This scaling is nevertheless only valid under the assumption that the gas streamlines follow a pure Keplerian shear flow. In reality, the gravity of the planet will bend the streamlines, turning those streamlines with small impact parameter into horseshoe flows \citep{Ormel2013}. The hydrodynamical simulations of \cite{Popovas+etal2018} demonstrated that pebbles of all sizes penetrate the outer regions of the Hill sphere along bent gas streamlines. The smallest pebbles are then sorted away along the horseshoe streamlines and along more distant streamlines that pass relatively unperturbed through the Hill sphere. Larger pebbles detach from a wider interval of gas streamlines and sediment towards the protoplanet, encountering the gas envelope approximately at the distance of the Bondi radius, defined here as $R_{\rm B} = G M /c_{\rm s}^2$ where $c_{\rm s}$ is the sound speed of the gas in the protoplanetary disc.

\subsection{Recycling flows}

Recycling flows are characterised by streamlines that penetrate into the Hill
sphere and leave back to the protoplanetary disc again \citep{Alibert2017}. Thus
both horseshoe flows and the perturbed Keplerian shear can be considered recycling flows \citep{LambrechtsLega2017,KurokawaTanigawa2018}.
The recycling flows
nevertheless penetrate only to the Bondi radius, unless the envelope is nearly adiabatic
\citep{LambrechtsLega2017,Popovas+etal2018}. The reason why the Bondi radius marks the maximum penetration of the recycling flows is that the entropy can only be significantly reduced compared to the disc value interior of the Bondi radius \citep{Rafikov2006,PisoYoudin2014} -- and buoyancy effects due to entropy gradients prevent the penetration of the recycling flows \citep{KurokawaTanigawa2018}. We therefore assume that any
pebbles or pebble fragments that make it below the Bondi radius are protected
from the recycling flows, unless they are pushed out of the Bondi radius again
with rising convective gas plumes.

\section{Pebble evolution without convective gas motion}
\label{s:noconvection}

We analyze in this section the characteristic pebble size while ignoring the
convective gas motion. We use this approach to put into context the results
including convection presented in the following section.

\subsection{Sedimentation speed}

\begin{figure}
  \begin{center}
    \includegraphics[width=0.90\linewidth]{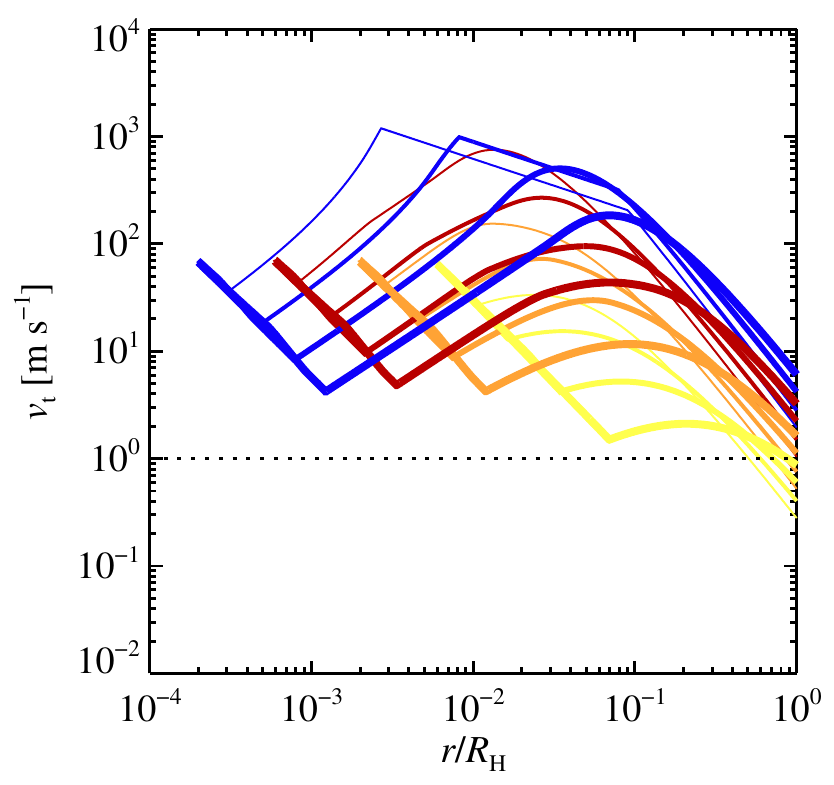}
  \end{center}
  \caption{The terminal velocity of the pebbles as a function of the distance
  from the protoplanet. We mark the critical speed for fragmentation in
  collisions between equal-sized pebbles (1 m/s). The general behaviour
  starting at the Hill radius is (a) an increase in the terminal velocity as the
  gravity increases, (b) for the low-mass protoplanets a change of slope where
  the free fall speed is lower than the terminal velocity, (c) a reduced
  sedimentation speed where the gas density increases closer to the protoplanet,
  and (d) for the high-mass protoplanets an increase in the terminal velocity
  starting at the transition between Epstein drag and Stokes drag. The highest
  sedimentation speeds are obtained for low-mass protoplanets far from the star,
  since these protoplanets have the lowest gas density in the envelope.}
  \label{f:terminal_velocity}
\end{figure}

Pebbles within the pebble capture radius sediment towards the protoplanet at the
terminal speed
\begin{equation}
  v_{\rm t} = \tau_{\rm f} \frac{G M}{r^2} \, .
\end{equation}
Here $\tau_{\rm f}$ is the friction time of the pebbles and $r$ is the distance
from the protoplanet. The collision speed of pebbles is given by the
differential sedimentation speed and the speed from the turbulent convection
(which we ignore in this section). The shear speed does not contribute
to the collision speed, since dust, pebbles and gas follow the same streamlines
outside of the Bondi radius of the protoplanet. The terminal velocity of the
pebbles is shown in Figure \ref{f:terminal_velocity}. We indicate also the
critical speed for fragmentation in collisions between equal-sized pebbles (1
m/s) with a dotted line \citep{Guettler+etal2010,Steinpilz+etal2019}. The
highest sedimentation speeds are obtained for low-mass protoplanets far from the
star, while the high gas density closer to the star acts to substantially slow
down the pebbles.
\begin{figure}
  \begin{center}
    \includegraphics[width=0.90\linewidth]{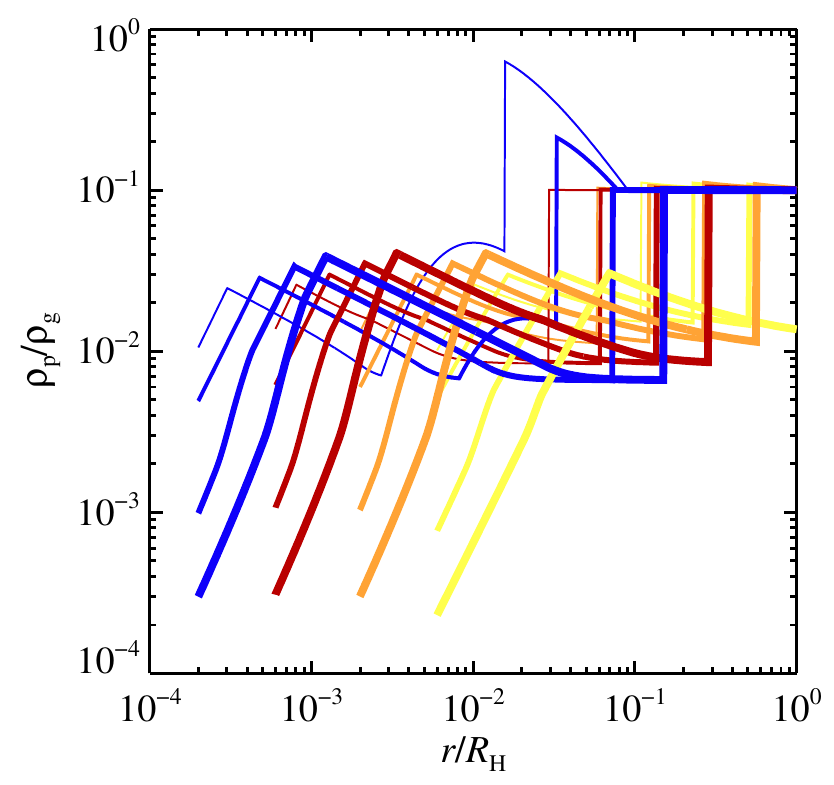}
  \end{center}
  \caption{The mass density of the pebbles relative to the gas as a function of
  the distance from the protoplanet. We assumed here that the pebbles fall at
  their terminal velocity (or the free fall speed when that is slower than the
  terminal velocity; this gives rise to the increased pebble-to-gas ratio
  outside of the Bondi radius for the two lowest-mass protoplanets furthest from
  the star) and that the mass flux is independent of the distance from the
  protoplanet. The pebble-to-gas ratio is constructed to match an outer boundary
  value of $0.1$. The sudden fall in the pebble-to-gas ratio marks the
  transition to spherical symmetry inside of the Bondi radius. This is followed
  by an increased pebble-to-gas ratio as the temperature increases, before the
  transition from Epstein to Stokes drag leads to a faster sedimentation speed
  and a decreased pebble-to-gas ratio near the surface of the protoplanet.}
  \label{f:dust_to_gas_ratio}
\end{figure}

\subsection{Pebble-to-gas ratio}

The density of the accreted pebbles relative to the gas can be calculated from
the assumption that the mass flux of pebbles is constant, starting from the
pebble accretion radius. We will furthermore assume that the pebble density is
spherically symmetric inside of the Bondi radius, yielding the expression
\begin{equation}
  \dot{M} \equiv \frac{M}{\tau} = 4 \pi r^2 \rho_{\rm p} v_{\rm t} = {\rm
  constant} \, ,
  \label{eq:Mdot}
\end{equation}
where $\dot{M}$ is the mass accretion rate, $M$ is the protoplanet
mass, $\tau$ is the accretion time-scale and $\rho_{\rm p}$ is the density of
the particles. The pebble-to-gas mass ratio is therefore
\begin{equation}
  \epsilon = \frac{\rho_{\rm p}}{\rho_{\rm g}} = \frac{\dot{M}}{4 \pi r^2 v_{\rm
  t} \rho_{\rm g}} = \frac{\dot{M} c_{\rm s}}{4 \pi G M R \rho_\bullet} \, .
  \label{eq:eps}
\end{equation}
Here the last step is valid for Epstein drag with friction time $\tau_{\rm f} =
R \rho_\bullet/(c_{\rm s} \rho_{\rm g})$, where $\rho_\bullet$ is the material
density of the particles, and terminal velocity $v_{\rm t} = \tau_{\rm f} G
M/r^2$ \citep[see][for discussions of the different drag force regimes
relevant for protoplanetary
discs]{Whipple1972,Weidenschilling1977,Johansen+etal2014}.
The pebble-to-gas ratio therefore remains constant in the isothermal regions of
the envelope and rises slowly proportional to the increase in sound speed in the
deeper regions.

Outside of the Bondi radius we cannot assume spherical symmetry. We
parameterize the degree of spherical symmetry by calculating the accretion
factor
\begin{equation}
  f_{\rm sphere} = \frac{\rho_{\rm p, mid}}{\rho_{\rm p,sphere}} \, .
  \label{eq:fsphere}
\end{equation}
Here $\rho_{\rm p,mid}$ is the mid-plane pebble density for the undisturbed flow
and $\rho_{\rm p,sphere}$ is the mass density obtained from the spherically
symmetric assumption, equation (\ref{eq:Mdot}). We assume that the mid-plane has
a pebble-to-gas ratio $\rho_{\rm p,mid}/\rho_{\rm g}=0.1$ outside of the Hill
radius. We multiply the pebble density obtained by assumption of spherical
symmetry by this accretion factor, evaluated at the Hill radius, to
maintain the coverage of the unit sphere at the accretion radius all the way
towards the Bondi radius. 

We show the calculated pebble-to-gas mass ratio for our models in Figure
\ref{f:dust_to_gas_ratio}. The pebble-to-gas ratio stays below a few percent
throughout most of the envelope. Thus our assumption in making Figure
\ref{f:envelope} that the opacity is given by micron-sized grains at 1\% mass
loading is not valid, since millimeter-sized particles yield a (geometric)
opacity that is orders of magnitude lower than micron-sized particles.
\cite{Ali-DibThompson2019} showed that small particles will pile up in the
envelope until the envelope has high enough opacity to become convective.  We
therefore consider in the main paper only the two extremes where the envelope is
either fully radiative or fully convective. We do not explore here further the
feedback between opacity and temperature, since we will demonstrate in Section
\ref{s:convection} that in the more realistic case where the large-scale
convective motion of the gas is included, the speed of both pebbles, pebble
fragments and dust is set mainly by the speed of the gas -- and hence pebble
fragments and dust cannot pile up in the envelope once the envelope becomes
convective.

\subsection{Mean collision distance}

\begin{figure}
  \begin{center}
    \includegraphics[width=0.90\linewidth]{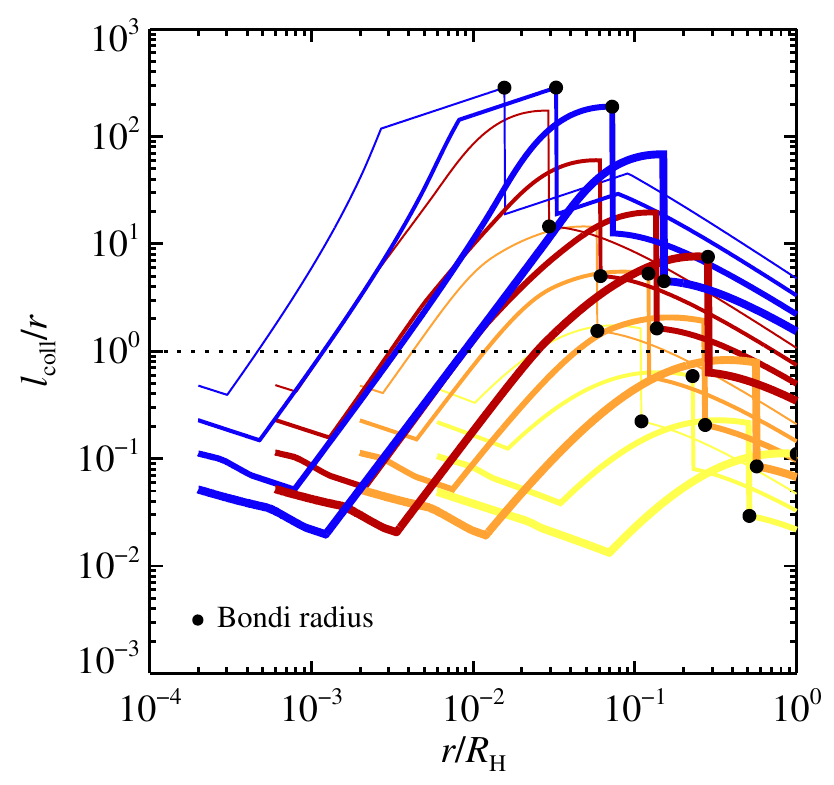}
  \end{center}
  \caption{The mean collision distance of the pebbles, relative to the distance
  to the protoplanet, as a function of distance from the protoplanet. The
  collision distance increases with distance from the star and is generally
  longer than the distance to the protoplanet in the regions of the Hill sphere
  and the Bondi sphere where the sedimentation speed is high.}
  \label{f:mean_free_path}
\end{figure}

High relative speeds do not necessarily imply fragmenting collisions, since the
pebbles must also have time to collide on the way towards the protoplanet. The
mean collision distance can be calculated from the pebble number density $n_{\rm
p}$ and cross section $\sigma_{\rm p}$ as
\begin{equation}
  \ell_{\rm coll} = \frac{1}{n_{\rm p} \sigma_{\rm p}} = \frac{(4/3) R
  \rho_\bullet}{\rho_{\rm p}} \, .
\end{equation}
We assumed here spherical pebbles with a constant internal density
$\rho_\bullet$. We calculate the pebble density $\rho_{\rm p}$ from equation
(\ref{eq:Mdot}) and equation (\ref{eq:fsphere}). The mean collision distance is
shown in Figure \ref{f:mean_free_path}, normalised by the distance to the
protoplanet. The mean collision distance is generally longer than the distance
to the protoplanet outside of the Bondi radius when the protoplanet is far from
the star. These are also the regions of highest terminal velocity (compare to
Figure \ref{f:terminal_velocity}). Protoplanets closer to the star are
collisional also outside of the Bondi radius, but the sedimentation speeds are
relative modest, in the 1--10 m/s range, at the higher gas densities closer to
the star.

\subsection{Erosion distance}

We calculated the mean collision distance above based on pebble-pebble
collisions. However, the envelope may contain a significant population of small
grains as well. The maximum projectile size giving rise to erosion of a pebble,
at the relative speed is $v$, was measured in \cite{Schraepler+etal2018} to be
\begin{equation}
  R_{\rm eros} = 2 \times 10^{-5}\,{\rm m}\,\left( \frac{v}{15\,{\rm m\,s^{-1}}}
  \right)^{1.62} \, .
  \label{eq:aeros}
\end{equation}
Thus any impactor smaller than 20 $\mu{\rm m}$ will erode the target when the
collision speed is 15 m/s and the limit rises to $0.4$ mm at a collision speed
of 100 m/s. The terminal velocity of the pebbles reaches high enough values
high up in the envelope for collisions with smaller dust aggregates to be
erosive. Large projectiles are less efficient than small projectiles at eroding
dust aggregate pebbles. The mass loss observed in the experiments of
\cite{Schraepler+etal2018} was fitted as
\begin{equation}
  \frac{\Delta m_{\rm p}}{m_{\rm proj}} = \left( \frac{v}{15\,{\rm m\,s^{-1}}}
  \right) \left( \frac{R_{\rm proj}}{2 \times 10^{-5}\,{\rm m}} \right)^{-0.62}
  \, .
  \label{eq:dmp}
\end{equation}
This expression gives an erosion efficiency of 42 (200) for micron-sized grains
impacting at 100 (500) m/s and 2.5 (12.3) for projectile grains of 100 microns
in size. The erosion distance is given by
\begin{equation}
  \ell_{\rm eros} = \frac{m_{\rm p}/(f_{\rm proj} m_{\rm proj})}{n_{\rm proj}
  \sigma_{\rm p}} = \frac{m_{\rm p}}{f_{\rm proj} \rho_{\rm proj} \sigma_{\rm
  p}} \, ,
\end{equation}
where $m_{\rm p}$ and $\sigma_{\rm p}$ are the mass and cross section of the
pebble, $m_{\rm proj}$ is the mass of the projectile, $\rho_{\rm proj}$ and
$n_{\rm proj}$ are the number density and mass density of the eroding grains and
$f_{\rm proj} = \Delta m_{\rm p}/m_{\rm proj}-1$ is the erosion efficiency. The
erosion distance can now be written in terms of the pebble-pebble collision
distance as
\begin{equation}
  \ell_{\rm eros} = \frac{\ell_{\rm coll}}{f_{\rm proj} (\rho_{\rm
  proj}/\rho_{\rm p})} \, ,
\end{equation}
where $\rho_{\rm p}$ is the mass density of pebbles in the gas. Inserting the
erosion efficiency from \cite{Schraepler+etal2018} and adopting the size
distribution ${\rm d}n_{\rm proj}/{\rm d}R \propto R^{-q}$ we arrive at
\begin{equation}
  \frac{\ell_{\rm eros}}{\ell_{\rm coll}} =
  \left(\frac{R_{\rm proj}}{R_{\rm p}}\right)^{q-4}
  \left(\frac{R_{\rm proj}}{20\,{\rm \mu m}}\right)^{0.62}
  \left(\frac{15\,{\rm m\,s^{-1}}}{v}\right)\, .
\end{equation}
Here the exponent $q - 4$ comes from multiplication by mass (scaling as $R^3$)
and averaging over a logarithmic mass interval (by multiplying by an additional
factor $R$) -- the relevant integral is shown in equation (\ref{eq:K}). We
see that for $q=3$, a size distribution with equal surface area
in all particle sizes, the erosion distance by micron-sized grains is longer
than the mean pebble-pebble collision distance for speeds slower than 2,300 m/s
(which is well outside of the range of erosion experiments). For grains of 100
microns the limiting speed is 400 m/s. Adopting instead a size distribution
power law of $q=3.5$, micron-sized grains erode as efficiently as pebble-pebble
collisions at a speed of 70 m/s, while grains of 100 microns erode as well as
pebbles at a speed of 130 m/s. We thus conclude that erosion by collisions with
smaller grains can at most be as efficient as pebble-pebble collisions at
destroying the pebbles and only at the highest sedimentation speeds experienced
by the captured pebbles. Such high speeds occur mainly for our protoplanets
growing at 30 AU, due to the low gas density there. Also, the mean pebble-pebble
collision distance is typically 10 or more times the distance to the protoplanet
where such high sedimentation speeds occur. We ignored in this analysis the
possibility that the smaller dust aggregates could be very fluffy and would have
a much more limited erosion capability \citep{Seizinger+etal2013}.

This analysis implicitly assumed that the density of the projectiles (dust and
pebble fragments) is similar to the density of the pebbles. This is a reasonable
assumption in the initial capture process where pebbles of many sizes move along bent gas streamlines through the Hill radius. Inside of the Bondi radius, dust and fragments may
nevertheless pile up to very high densities. We consider the effect of such pile
ups on the erosion of pebbles in Section \ref{s:coagulation}.

\subsection{Fragmentation-limited pebble sizes (sedimentation)}

\begin{figure}
  \begin{center}
    \includegraphics[width=0.90\linewidth]{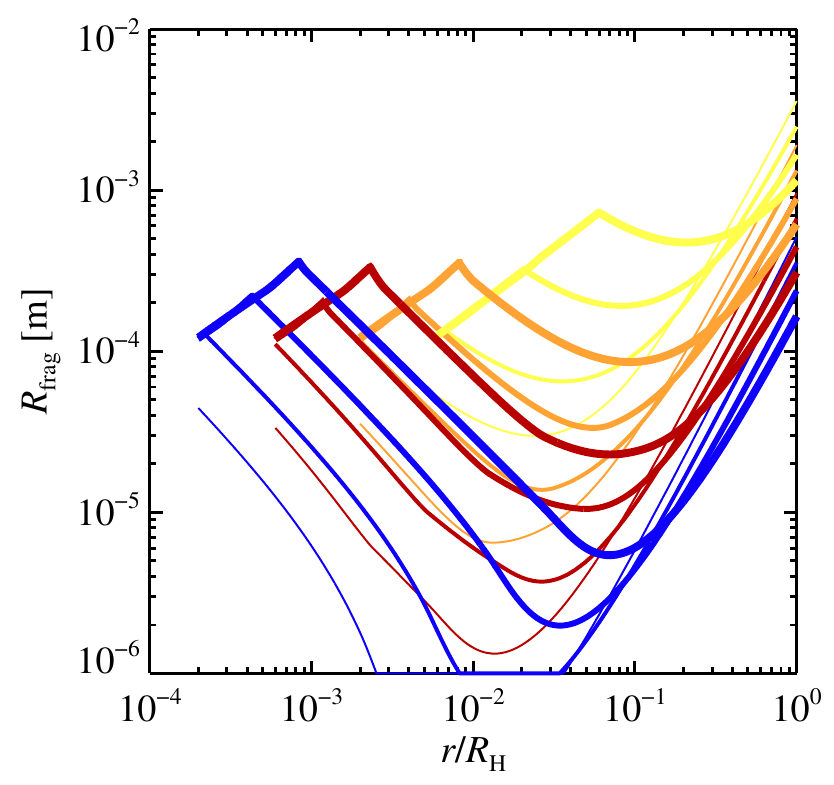}
  \end{center}
  \caption{Fragmentation-limited particle size, assuming a worst-scenario where
  the critical fragmentation speed is 1 m/s and the pebble-pebble collision
  speed is equal to the sedimentation speed, as a function of the distance from
  the protoplanet. The pebbles regrow to several hundred microns in size as they
  approach the lowest regions of the envelope where the gas density is high and
  the sedimentation speed is low.}
  \label{f:fragmentation_limit_settling}
\end{figure}

Collisions between equal-sized pebbles are expected to be destructive when the
collision speeds are higher than 1 m/s \citep{Guettler+etal2010}. This is a
well-established threshold for silicate dust aggregates and the same limit
likely applies to icy pebbles at low temperatures as well
\citep{MusiolikWurm2019}. Obtaining the fragmentation-limited particle size from
the sedimentation speed is nevertheless problematic, since equal-sized particles
formally have zero relative speed and collisions with particles of 100 microns
in size have an erosion threshold of 40 m/s (see equation \ref{eq:aeros}). The
threshold falls to 2--3 m/s for micron-sized projectiles
\citep{Schraepler+etal2018}. Taking the worst-case scenario where pebbles can
have a maximum sedimentation speed of 1 m/s, we plot in Figure
\ref{f:fragmentation_limit_settling} the fragmentation-limited pebble size.  The
size falls steadily with depth, until reaching the regions of high gas density
close to the protoplanet where the pebbles re-coagulate to sizes between 100
microns and 1 millimeter. The minimum pebble size is larger than 10 $\mu{\rm m}$
in the collisional regions of the envelope.

Full solutions to the coagulation equation presented in
\cite{Schraepler+etal2018} showed that particles in the protoplanetary disc grow
to dm sizes where they move with approximately 10 m/s through the gas. The
stalling of the growth at 10 m/s in \cite{Schraepler+etal2018} seems to be the
result of a balance between eroding collisions with small dust grains and
sticking collisions with medium-sized dust aggregates. In Section
\ref{s:coagulation} we also demonstrate that time-dependent solutions to the
erosion and growth of the pebbles yield systematically larger pebble sizes than
in Figure \ref{f:fragmentation_limit_settling} (compare to the right-side
plot of Figure \ref{f:size_sedimentation}).

\section{Pebble evolution including convective gas motion}
\label{s:convection}

The luminosity of protoplanets accreting pebbles is in many cases large enough that the energy must be transported through the envelope by convection. When present, convective motions will have a large influence on the dynamics and collision speeds of the pebbles in the envelope. We acknowledge that for some of the cases considered here, under given assumptions the results indicate that there may not be a sufficient presence of small dust to provide the opacity necessary to trigger convection. \cite{Ali-DibThompson2019} derived a criterion for the dust-to-gas ratio needed to drive convection (their equation 33). In Appendix B we therefore present additional numerical experiments where we fix the density of pebble fragments at 1\% and 10\% of the gas density, to bracket the values found by  \cite{Ali-DibThompson2019}. While an increase in the fragment density leads to a reduction in the pebble size reached at the planetary surfaces, we find qualitatively similar results when calculating the fragment density from the local pebble density (as we do in this section) and when considering a fixed fragment fraction relative to the gas (as in Appendix B).

\subsection{Convective speed}

We use here a simple mixing length estimate of the luminosity transported by
convection \citep{Ali-DibThompson2019},
\begin{equation}
  L_{\rm c} = 4 \pi r^2 (0.36 \alpha_{\rm mix}) \rho_{\rm g} v_{\rm c}^3 \, .
\end{equation}
Here $\alpha_{\rm mix}$ is a mixing length coefficient that we take to be unity
and $v_{\rm c}$ is the characteristic speed of the convection cells. Setting
$L_{\rm c}$ equal to the luminosity of pebble accretion, $L_{\rm c} = G M
(M/\tau) / R$, where $\tau$ is assumed accretion time-scale and $R$ is the
radius of the solid protoplanet, we obtain an approximate value for $v_{\rm c}$.
We fix in the main paper $\tau=10^6\,{\rm yr}$ and note that the choice of
$\tau$ affects the gas densities in the envelope and hence the collision speeds.
In Appendix A we explore the effect of a lower value of $\tau$.

\begin{figure*}
  \begin{center}
    \includegraphics[width=0.45\linewidth]{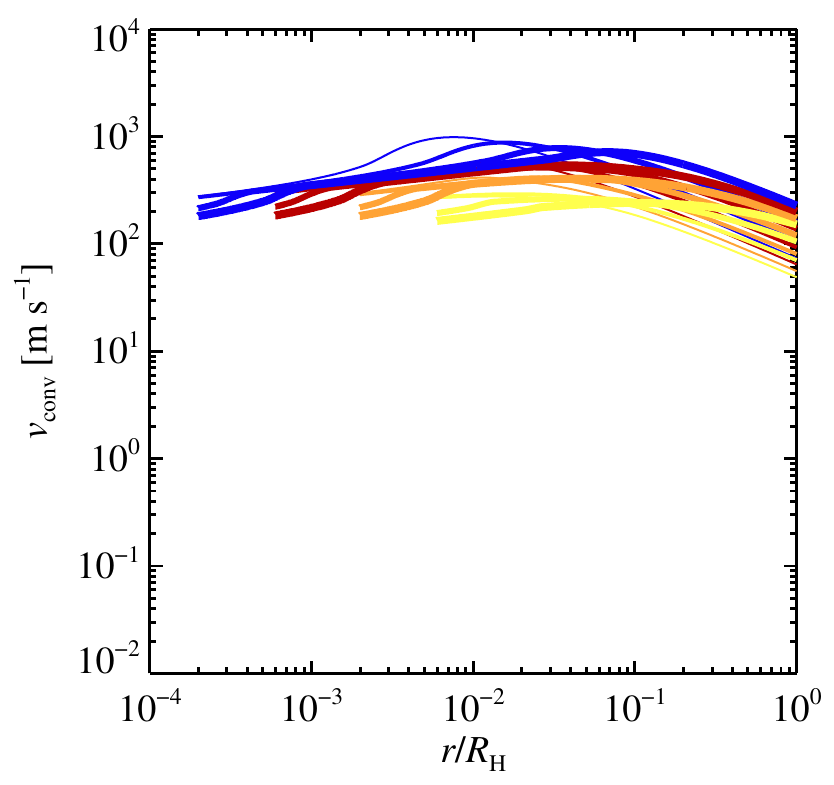}
    \,
    \includegraphics[width=0.45\linewidth]{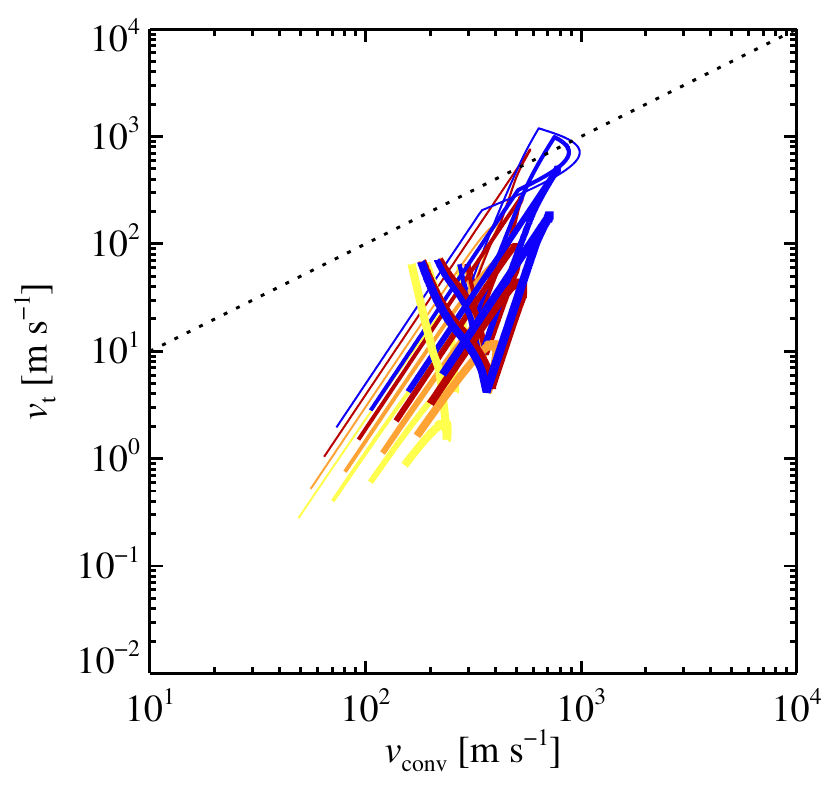}
  \end{center}
  \caption{The convection speed, from a mixing length expression based on the
  luminosity of the protoplanet, as a function of the distance from the
  protoplanet (left) and the terminal velocity as a function of the convection
  speed (right). The convective speed is generally larger than the sedimentation
  speed (the dotted line marks the equality between the convective speed and the
  terminal velocity). All particle sizes are thus strongly affected by transport
  with convection cells, as also seen in 3-D simulations in
  \cite{Popovas+etal2018} and \cite{Popovas+etal2019}. Pebbles can therefore
  only accreted to the protoplanet with the downwelling cold gas, if the
  envelope is convective.}
  \label{f:convection_speed}
\end{figure*}

We plot the convective speed estimates in Figure \ref{f:convection_speed}. The estimates generally lie between 100 m/s and 1,000 m/s, and are thus always higher than the local sedimentation speed. The dynamics of the pebbles is therefore dominated by transport with the convective gas rather than by sedimentation. 

We point out here that, contrary to what is normally assumed when deriving mixing length estimates, the ratio of the local scale height to the radius is not small in the case of a protoplanet envelope. Hence the scale of the dominating convective motions can even be comparable to the system scale (cf. \cite{Popovas+etal2018}), and should thus not be thought of as small-scale turbulent motions.

The capture of pebbles is ultimately determined from the
streamlines that enter the Hill radius---smaller pebbles, whose paths deviate less from their streamlines, must enter on streamlines that reach closer to the planet, to avoid being transported past the protoplanet
\citep{Ormel2013}. The
streamline interval for accreted pebbles thus narrows for decreasing pebble size.
When convection is present the pebbles are transported faster towards the
protoplanet with downwelling cold gas, but also visit for a correspondingly
shorter time, and as observed in the 3-D hydrodynamical simulations of
\cite{Popovas+etal2018} and \cite{Popovas+etal2019} the accretion rates are
therefore not strongly affected when constant particle size is assumed. 
How the value of the limiting fragment size
that can detach at the base of the convection flow depends on the protoplanet's
mass and distance to the star is nevertheless relatively poorly understood,
since the simulations of \cite{Popovas+etal2019} focused on the 1-1.6 AU region, and did not include fragmentation, coagulation, and sublimation.

\subsection{Pebble-to-gas ratio (convection)}

\begin{figure}
  \begin{center}
    \includegraphics[width=0.90\linewidth]{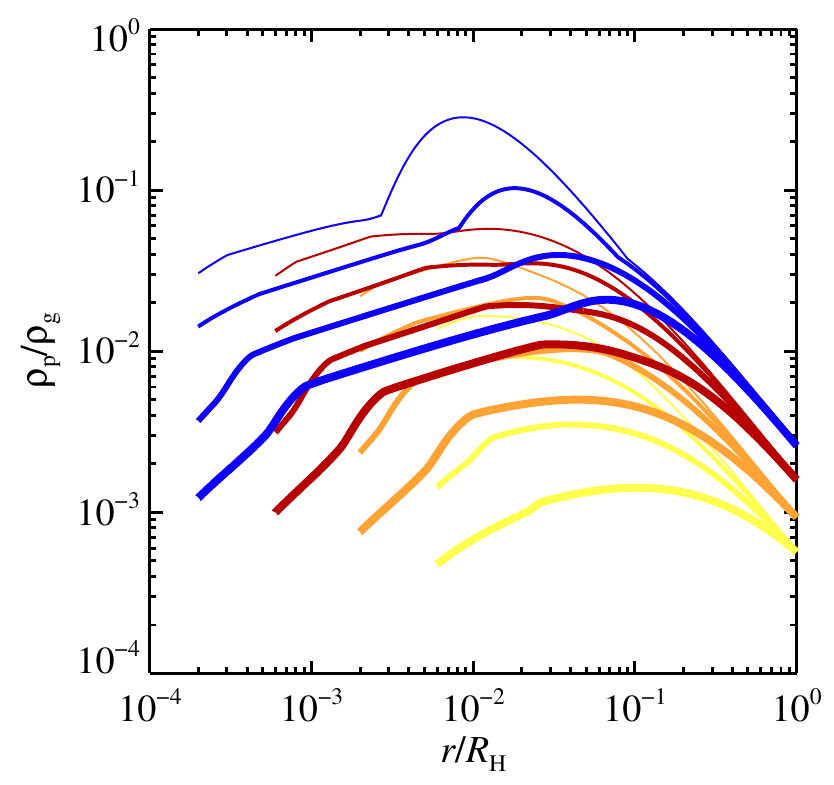}
  \end{center}
  \caption{The pebble-to-gas ratio in the convective case. We assume here that
  the speed of the pebbles is given by the downwards speed of the convective
  cells, starting already at the pebble accretion radius due to convective
  overshoot. We furthermore assume that the mass flux is not spherically
  symmetric inside of the Bondi radius, since the pebbles are transported
  towards the surface of the protoplanet along downwelling cold flows. These two
  effects partially cancel each other so that the pebble-to-gas ratio is
  relatively similar to the non-convective case shown in Figure
  \ref{f:dust_to_gas_ratio}.}
  \label{f:dust_to_gas_ratio_convection}
\end{figure}

The high speeds of the convective cells imply that pebbles are transported
towards the protoplanet's surface along downwelling cold flows with a low
ambient
pebble density. However, we can no longer assume that the transport is
spherically symmetric within the Bondi radius, as the downwelling cold flows
transport pebbles from the accretion radius all the way down to the surface in a
single turn-over time-scale \citep{Popovas+etal2019}. We therefore calculate the
pebble-to-gas ratio inside of the Bondi radius in the convective case including
the factor $f_{\rm sphere}$ from equation (\ref{eq:fsphere}), which maintains
the degree of spherical symmetry of the flow defined at the Hill
radius. The resulting pebble-to-gas ratio is shown in Figure
\ref{f:dust_to_gas_ratio_convection}. The increased speed of the gas flow is
partially cancelled by the non-spherical-symmetry of the pebble component and
hence the pebble-to-gas ratio appears quite similar to the non-convective case
shown in Figure \ref{f:dust_to_gas_ratio}. The high gas speed nevertheless means
that the pebbles have less time to collide on the way to the surface of the
protoplanet.

\subsection{Convective collision speeds}

\begin{figure}
  \begin{center}
    \includegraphics[width=0.90\linewidth]{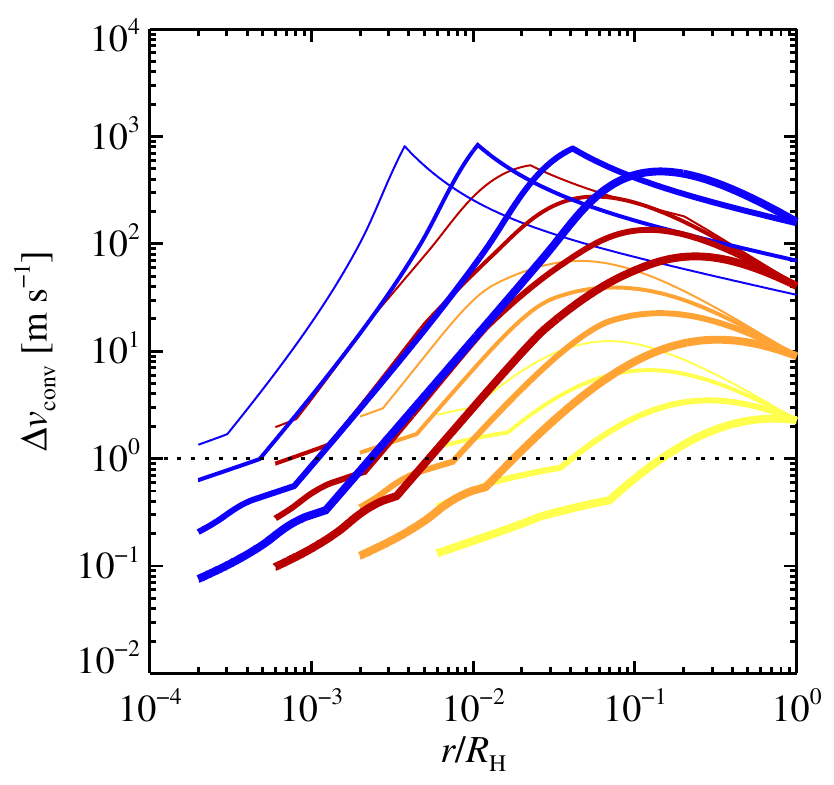}
  \end{center}
  \caption{The convective collision speed between millimeter-sized pebbles as a
  function of the distance from the protoplanet. We mark the critical
  fragmentation speed for equal-sized
  particles (1 m/s). The collisional regions of the envelope generally have
  convective collision speeds slower than 100 m/s.}
  \label{f:convection_collision_speed}
\end{figure}

The convective gas motions also induce collisions between the particles.
Inspired by calculations of the collision speed in generalized protoplanetary
disc turbulence \citep{OrmelCuzzi2007}, we take the collision speed $\Delta
v_{\rm c}$ to be
\begin{equation}
  \Delta v_{\rm c} = \sqrt{{\rm St}_{\rm c}} v_{\rm c} \, .
  \label{eq:dvc}
\end{equation} Here ${\rm St}_{\rm c} = \omega_{\rm c} \tau_{\rm f}$ is the
large-scale Stokes number of the pebbles and $\omega_{\rm c}$ is the frequency
of the large-scale turbulent convection cells. We use the simple expression
$\omega_{\rm c} = v_{\rm c}/R_{\rm B}$ for convection cells moving at speed
$v_{\rm c}$ over the Bondi radius $R_{\rm B}$. The convective collision speeds
between millimeter-sized pebbles are shown in Figure
\ref{f:convection_collision_speed}. The collision speed reaches several hundred
m/s for protoplanets residing far from the star. But these regions of high
convective collision speeds have long collision distances as we also saw for the
collisions driven by the sedimentation (compare to Figure
\ref{f:mean_free_path}). The collisional regions of the envelope have convective
collision speeds below 100 m/s.

\subsection{Fragmentation-limited pebble sizes (convection)}

\begin{figure}
  \begin{center}
    Convection only:
    \includegraphics[width=0.90\linewidth]{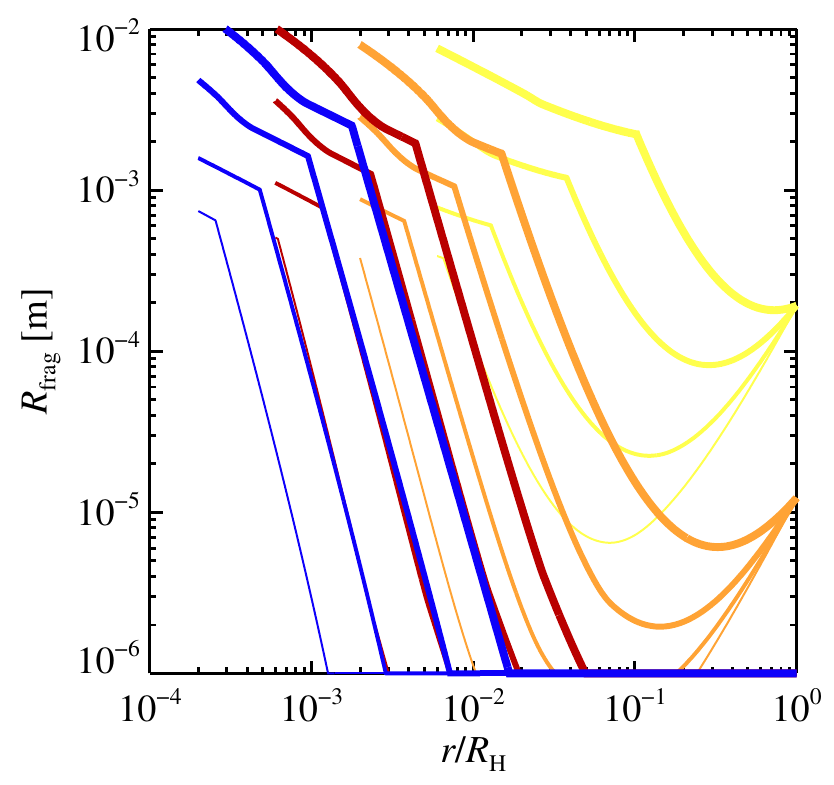}
    \\
    Convection and sedimentation:
    \includegraphics[width=0.90\linewidth]{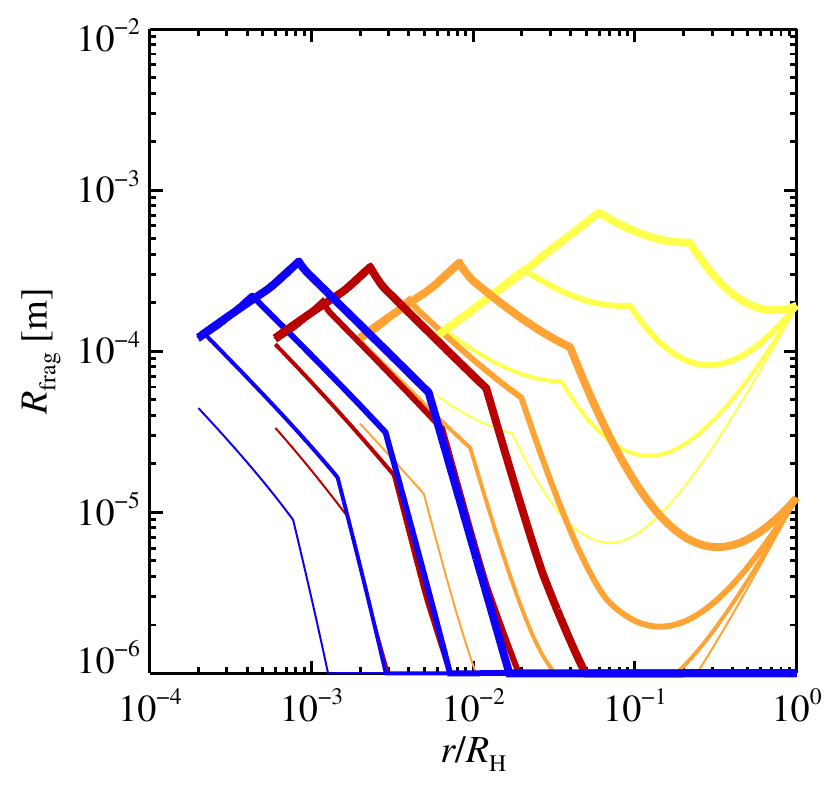}
  \end{center}
  \caption{Fragmentation-limited particle size assuming a critical fragmentation
  speed of 1 m/s and the relative speed from convection as a function of the
  distance from the protoplanet (top panel) and the fragmentation-limited
  particle sized resulting from adopting the smallest of the sizes resulting
  from turbulent    convection and sedimentation acting separately
  (bottom panel)}. The fragmentation-limited particle size reaches 1 micron or
  less at planetary distances between 3 AU and 30 AU. The regions of the
  envelope where the fragmentation-limited leads to monomer sizes are generally
  optically thin to collisions between pebbles, except for the most
  massive protoplanets (compare to Figure \ref{f:mean_free_path}). The pebbles
  nevertheless could recoagulate to sizes of 0.1-1 millimeter as they
  enter the regions of high density near the protoplanet, if there is time
  enough to grow to the fragmentation-limited size during the downwards
  transport with the gas.
  \label{f:fragmentation_limit_convection}
\end{figure}

The fragmentation-limited pebble size for convective collisions can be
calculated from
\begin{equation}
  {\rm St}_{\rm c} = \left( \frac{v_{\rm frag}}{v_{\rm c}} \right)^2 \, .
\end{equation}
Here we again take a conservative fragmentation threshold of $v_{\rm frag}=1$ m/s. The convective Stokes number can then be converted to a particle size when the gas density and temperature are known. The resulting particle sizes are shown in the upper panel of Figure \ref{f:fragmentation_limit_convection}. 
If convection is assumed to occur in the form of small scale turbulent motions it is much more efficient at fragmenting the pebbles than sedimentation, since the relative speed is only reduced as the square root of the particle size in the case of turbulence. Hence protoplanets at 10 AU and 30 AU have fragmentation-limited particle sizes smaller than the size of a monomer in the outer regions of the envelope. These regions are nevertheless optically thin to collisions. However, the fragmentation-limited pebble size increases further down in the envelope and reach millimeter sizes again close to the surface of the protoplanet. 

The lower panel of Figure \ref{f:fragmentation_limit_convection} shows the results of adopting the smallest of the particle sizes  from the cases with sedimentation and convection acting separately.  This is likely a worst case estimate, in that the large scale convection found in \cite{Popovas+etal2019} also would act to reduce the fragmentation in downwards directed plumes, by shortening the transport time of particles down to the neighborhood of the surface.

\section{Solving the pebble size evolution}
\label{s:coagulation}

The analysis of the fragmentation-limited pebble sizes in the previous two
sections ignored many aspects of pebble erosion and pebble growth as well as the
time-scale needed to reach the fragmentation-limited size. We therefore go 
a step further in analysing the size evolution of pebbles in the envelope by
solving for the time-dependent pebble size, taking into account erosion,
fragmentation and growth by mass transfer.

\subsection{Size distribution and fragment density}

We follow the pebble size $R_{\rm p}$ and assume that smaller particles are
present as a continuous size distribution with the number density ${\rm d}n/{\rm
d}R = K R^{-q}$ down to the smallest particle size $R_0=10^{-6}\,{\rm m}$ (we
fix $q=3.5$ here).  The size distribution is normalised to give the total dust
mass from equation (\ref{eq:Mdot}), with
\begin{equation}
  \int_{R_0}^{R_{\rm p}} K R^{3-q} {\rm d} R = \rho_{\rm p} \, .
  \label{eq:K}
\end{equation}
We make one of two assumptions about the sizes and total density of the particles: (i)
that the fragment size distribution extends to the pebble size and that the
total density of pebbles and fragments is given by the radial speed of the
pebbles or (ii) that the fragment size distribution extends only to the
fragmentation-limited size and that these fragments have the same flux as the
pebbles within the Bondi radius, with their density given by the speed at the
fragmentation-limited size. The first case is realistic if pebbles and fragments
enter a collisional equilibrium whereby the net downwards motion is set by the
largest objects. The second assumption is a worst-case scenario where the entire
pebble flux is converted into fragments that achieve a high spatial density due
to their slow sedimentation speed. We always assume assumption (i) to be the
case outside of the Bondi radius, since the shear flows and recycling flows
prevent the pile up of fragments in this region and additionally the pebble
accretion process itself intrinsically picks up large particles out of the shear flow and transport them towards the protoplanet, leaving behind the smaller particles. 
We additionally limit the fragment density to be less than or equal to the gas density inside
of the envelope, since any additional mass loading is dynamically unstable to
the formation of rapidly sedimenting Rayleigh-Taylor-like dust blobs
\citep{Lambrechts+etal2016,Capelo+etal2019}. We have done additional tests using
the fragmentation-limited approach also outside of the Bondi radius as well as
allowing pebble-to-gas ratios above unity and found only minor diffences in the
resulting pebble sizes.

\subsection{Pebble size evolution model}

We divide the particle size distribution into $N_{\rm spec}=100$ bins (or
species), spaced logarithmically so that ${\rm d} R \propto R$ when assigning
mass to each bin using the binned equivalent of equation (\ref{eq:K}). The size
of the particles in species $i$ is $R_i$, their mass is $m_i$ and their density
is $\rho_i$.

\begin{figure}
  \begin{center}
    \includegraphics[width=0.90\linewidth]{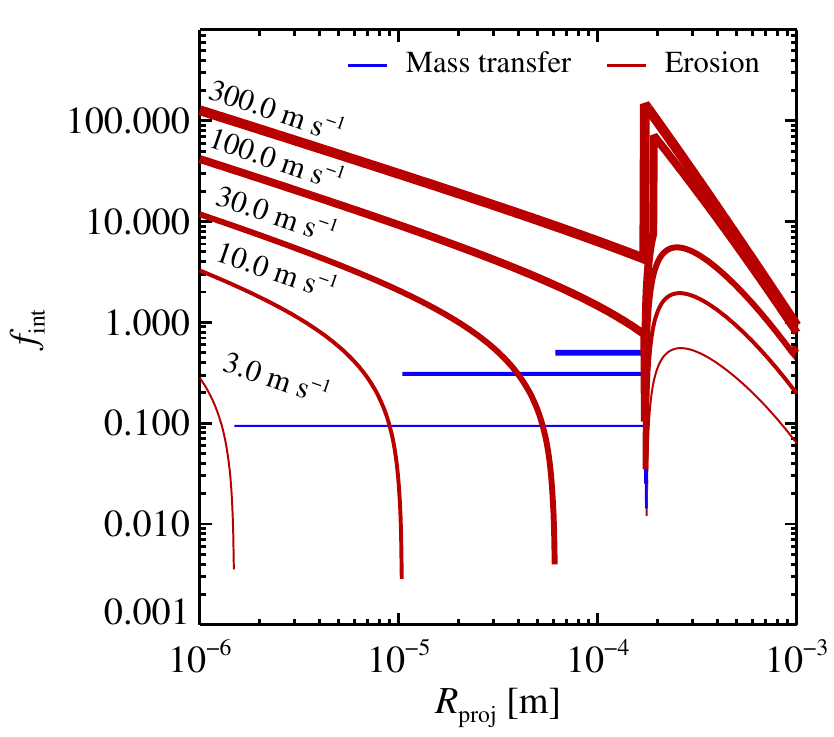}
  \end{center}
  \caption{The interaction fraction (i.e., the mass loss of the target relative
  to the projectile mass), as a function of the projectile size and shown for
  five separate collision speeds of $3.0$, $10.0$, $30.0$, $100.0$ and $300.0$ m
  s$^{-1}$. The erosion done by small projectiles was parameterised in
  \cite{Schraepler+etal2018}.  This is followed by a range of projectile sizes
  that transfer up to 50\% of their mass to the projectile, when the collision
  speed is lower than a threshold value around 50 m/s. Mass transfer transitions
  to fragmentation starting at a relative particle size of $5.83$
  \citep{BukhariSyed+etal2017}.}
  \label{f:erosion_factor}
\end{figure}

We change the mass of the pebbles by summing over the equation
\begin{equation}
  \dot{m}_{{\rm p}} = -\sum_i \pi (R_{\rm p}+R_i)^2 \Delta v_i \rho_i m_i
  f_{{\rm int},i} \, .
\end{equation}
Here the relative speed $\Delta v_i$ is given by either the relative
sedimentation speed (proportional to $R_{\rm p}-R_i$) or the combination of
relative sedimentation speed and the convective collision speed. The latter is
approximated as the collision speed of the pebbles, equation (\ref{eq:dvc}), as
this represents well the collision speed with smaller particles as well. The
interaction fraction $f_{{\rm int},i}$ can be either positive (erosion or
fragmentation) or negative (sticking or mass transfer). We follow
\cite{BukhariSyed+etal2017} and distinguish between the situation where the
projectile and the target have very dissimilar sizes ($R_{\rm p}/R_i \geq 5.83$)
and the situation where they have similar sizes ($R_{\rm p}/R_i < 5.83$). The
different collision outcomes are parameterised as
\begin{itemize}
  \item Erosion ($R_{\rm p}/R_i>5.83$). The eroded mass fraction relative to the
  projectile is given in \cite{Schraepler+etal2018}, see also our equation
  (\ref{eq:dmp}). We subtract one from the mass fraction to get the relative
  mass loss.
  \item Mass transfer ($R_{\rm p}/R_i > 5.83$). When the erosion coefficient of
  \cite{Schraepler+etal2018} becomes less than one, we use the power law fit
  from \cite{BukhariSyed+etal2017} to calculate the mass transfer coefficient
  as a function of the relative velocity, capped at a maximum of 0.5 at high
  speeds \citep{Wurm+etal2005}.
  \item Fragmentation ($R_{\rm p}/R_i < 5.83$). When the projectile is larger
  than $1/5.83$ times the target size, we allow the projectile to fragment the
  target. We use the power law fit to collision experiments provided in
  \cite{BukhariSyed+etal2017} to calculate the erosion factor for fragmenting
  collisions. We include the probability $p_{\rm sur} = 0.194 R_{\rm p}/R_i -
  0.13$ that the target survives the collision (i.e., is not fragmented) and
  instead experiences mass transfer from the projectile. Mass transfer is not
  allowed when the kinetic energy in the collision is higher than the energy
  needed to reduce the target mass by a factor two. The final erosion factor is
  calculated as a weighted average of mass transfer and erosion, $f_{\rm int} =
  -p_{\rm sur} f_{\rm mt} + (1-p_{\rm sur}) f_{\rm frag}$.
\end{itemize}
The interaction fraction $f_{\rm int}$ as a function of the projectile size is
shown for five different collision speeds in Figure \ref{f:erosion_factor}.

\subsection{Pebble size evolution without convection}

\begin{figure*}
  \begin{center}
    \includegraphics[width=0.90\linewidth]{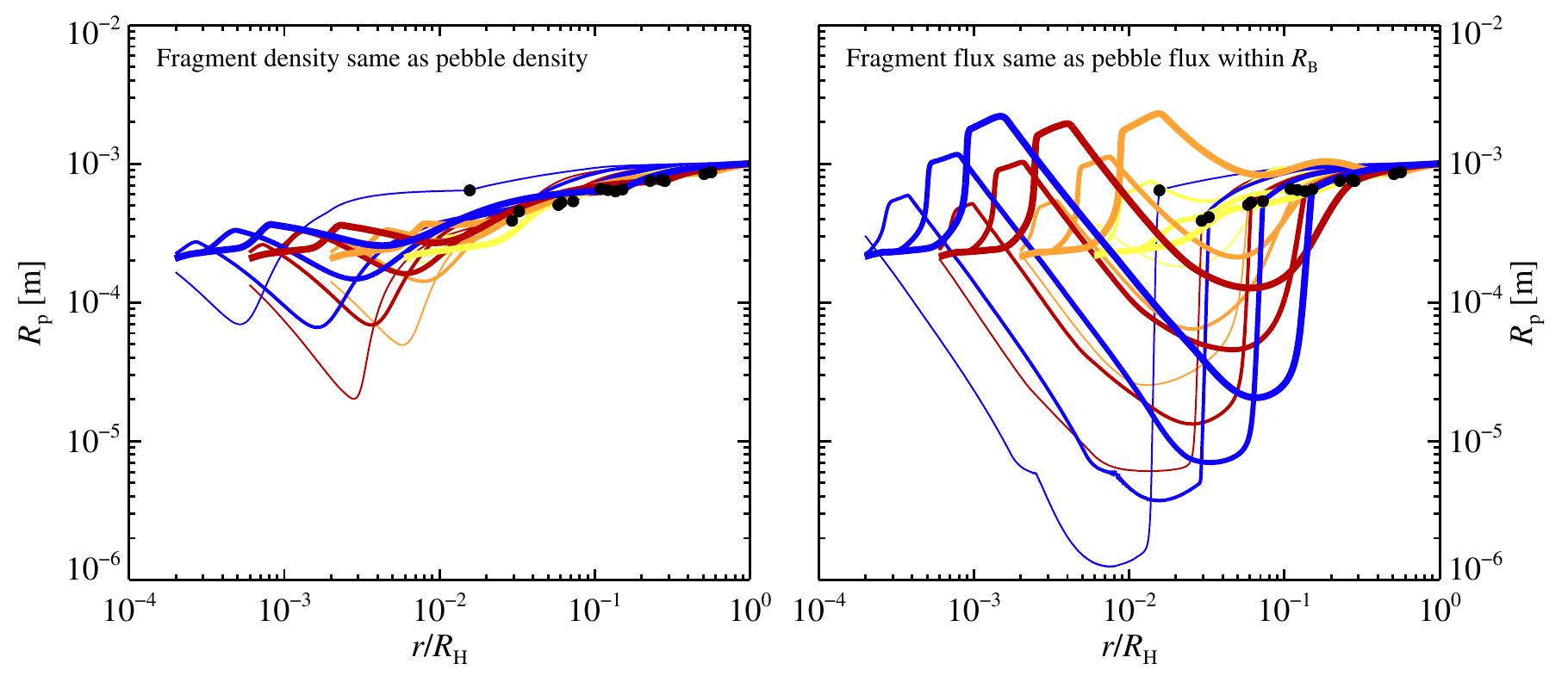}
  \end{center}
  \caption{The pebble size in models without convection, resulting from two
  assumptions about the density of the pebble fragments: (i) that their density
  is given by the radial speed of the pebbles (left) and (ii) that their density
  is given by the radial speed of the largest fragments, assumed to be at the
  fragmentation-limited size and with the same mass flux as the pebbles (right).
  We mark the Bondi radius with black dots; outside of the Bondi radius we
  assume that the fragment density is equal to the pebble density since
  particles cannot pile up in this region. The pebbles undergo
  only minor erosion outside of the Bondi radius; this contrasts with the small
  size of the fragmentation-limited particles seen in Figure
  \ref{f:fragmentation_limit_settling}. Within the Bondi radius we observe that
  model (i) sees only limited size changes, while model (ii) approaches the
  fragmentation-limited sizes. The pebbles are nevertheless a factor 2--3 times
  larger than the fragmentation limit inside of the Bondi radius.}
  \label{f:size_sedimentation}
\end{figure*}

In Figure \ref{f:size_sedimentation} we show the pebble size as a function of
the distance from the protoplanet. There is only little evolution in the pebble
size outside of the Bondi radius. The pebbles inside of the Bondi radius reach a
size near the fragmentation-limited expression near the surface of the
protoplanet (compare to Figure \ref{f:fragmentation_limit_settling}) for both
our assumptions about the fragment density -- and in the case where the fragment
mass flux equals the pebble mass flux the pebble sizes follow the
fragmentation-limited size even in the regions of the envelope that we
identified as optically thin to collisions. This is due to the high density of
pebble fragments in those regions from our assumption that the fragment flux is
equal to the pebble flux there.

\subsection{Pebble size evolution including convection}

\begin{figure*}
  \begin{center}
    \includegraphics[width=0.90\linewidth]{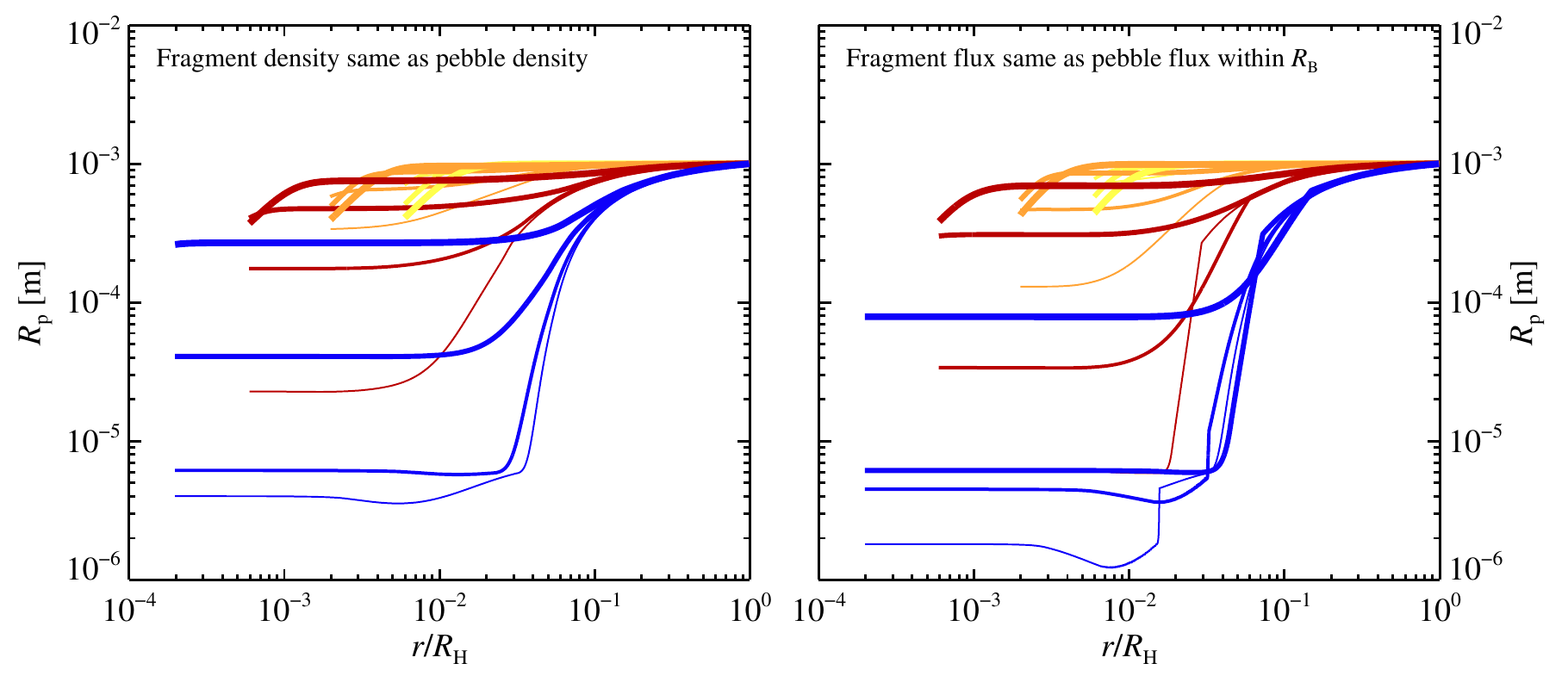}
  \end{center}
  \caption{The maximum pebble size resulting from the same two implementations
  of the size evolution model as in Figure \ref{f:size_sedimentation}, but here
  including the convective gas speed both for transport and for collisions. The
  downwards-moving cool gas flows are so fast that the pebbles do not have
  enough time to change sizes appreciably on their passage down to the
  protoplanet surface for the protoplanets at 1 and 3 AU. Protoplanets further out experience higher convective collision speeds due to the low
  gas density, especially when the protoplanets are of low mass. The pebbles are therefore ground down to sizes between a few and a few hundred microns.}
  \label{f:size_sedimentation_convection}
\end{figure*}

In Figure \ref{f:size_sedimentation_convection} we include the effect of
convection both for the speed of the pebbles as they move towards the
protoplanet (by adding to the sedimentation speed the downwards motion of the
cooling gas flows) and for the collision speed. 
The speed of the cool gas flows
are so high (see Figure \ref{f:convection_speed}) that the pebbles do not have
time to interact appreciably before they are accreted by the protoplanet, except
for the lowest-mass protoplanets far from the star. The difference between our
two fragment density prescriptions comes mainly here from the smaller size of
the fragments in case (ii), since the convective gas speed makes the pebble and
fragment density similar in both cases. In Appendix A we consider the evolution of pebbles at an earlier and a later evolution stage of the protoplanetary disc, respectively. The high gas densities in the protoplanetary disc early in the disc evolution reduces the collision speeds and leads to better pebble survival, while the opposite is the case for later evolution stages. \cite{Manara+etal2018} indeed suggested that the main phase of planetary growth happens at very early evolution stage and in these stages the pebbles also survive the passage down to the surface of the protoplanet better.

\section{Conclusions}
\label{s:conclusions}

In this paper we have presented calculations of pebble erosion, fragmentation
and regrowth inside of the envelope of an accreting protoplanet, covering a parameter range that includes both weakly and tightly coupled regimes (in terms of ratio of stopping time and collision time  to transit time at the Bondi radius). Pebbles reach
high relative speeds, up to of order 1,000 m/s, both by differential sedimentation and by acceleration in the convective gas. The speeds are highest for
low-mass protoplanets ($<1$ $M_{\rm E}$) far from the star ($a>3$ AU), at low protoplanetary disk accretion rates, while the
higher gas density obtained for higher accretion rates, and for high-mass protoplanets close to the star, reduces
the relative speeds. However, the regions of the envelope where
collision speeds are high are also typically optically thin to collisions.

We analysed the fragmentation-limited particle size, assuming a worst-case
scenario where the threshold speed for fragmentation is just 1 m/s. This leads
to very small pebble fragment sizes, down to micron sizes, in the regions of the envelope
where the relative speed is high. The increased gas density near the surface of
the protoplanet nevertheless allows regrowth to sizes up to a few hundred
microns.

We also implemented a simple size evolution scheme that keeps track of the
characteristic size of the pebbles as they move towards the surface of the
protoplanet. The size evolution depends strongly on our assumptions on the mass
density of the pebble fragments. Assuming that the fragment density follows the
pebble density, we find a modest decrease in the pebble size from the original
millimeter sizes to a few hundred microns near the surface of the protoplanet.
Allowing instead the fragments to pile up in the envelope, the pebble size
follows closely the fragmentation-limited size throughout the Bondi sphere. This
case seems to be most similar to the model of \cite{Ali-DibThompson2019}.

However, we have a significantly different approach to convection from
\cite{Ali-DibThompson2019}. Based on hydrodynamical simulations of accreting
protoplanets \citep{Popovas+etal2019}, we assume that the convective flow consists of large-scale flows
that extend from the protoplanet's surface all the way to the Bondi radius and
that the convective overshoot is significant enough to reach the pebble
accretion radius. Hence the pebbles and the fragments are transported down towards the protoplanet at speeds in the range 100-1,000 m/s, to near-stagnation regions close to the planetary surface, from where they can hence sediment down to the planet more easily.
This approach
precludes the pile up of pebble fragments in the envelope and additionally
lowers the residence time of the pebbles in the envelope, so that the erosion
effect is lowered. We observe then that the pebbles are transported to the
surface of the protoplanet at sizes between 100 microns and 1 millimeter.
Low-mass protoplanets residing far from the star ($>30$ AU) nevertheless still
experience significant pebble erosion due to the low gas densities in their
envelopes.

We have for simplicity ignored several physical effects in the model. We assumed
the opacity to be a constant, irrespective of the pile up of pebble fragments in
the envelope. The feedback from the opacity on the temperature structure was
included by \cite{Ali-DibThompson2019}, who concluded that the increased opacity
leads to a large-scale expelling of dust grains from the envelope. Their
approach to convection is nevertheless very different from ours, in that it was
assumed that convection can be treated as a purely diffusive process. We believe
that a more realistic view is to consider convection to be a large-scale
circulation that transports pebbles and fragments down to the protoplanet's
surface, while upwelling hot flows work to remove remaining dust and
fragments back to the protoplanetary disc.

We also ignored the effect of ice lines. Ice lines must be located interior of
the Bondi radius, since the temperature of the envelope only starts to rise
interior of the Bondi radius. Sublimation of ices (such as H$_2$O, CO$_2$ and
CO) would lead to a restructuring of the dust aggregate pebbles, but we assume
that the sublimation process is so slow (the sublimation time-scale is on the
order of several days or even weeks) that loss of volatiles does not lead to a
monomerisation of the aggregates. The silicate sublimation front at 2,000 K
would under all circumstances destroy pebbles and dust close to the surface of
massive protoplanets; however this sublimation is associated with a radiative
zone, due to the strong decrease in opacity there (see Figure 1), which would
separate the silicate vapour from the bulk envelope. In addition we ignored the
latent heat from sublimation and deposition
\citep{Brouwers+etal2018,Brouwers+etal2019,BrouwersOrmel2020}, which may also be
important for the thermodynamics of the envelope.

From our study we conclude that (i) if the gas envelope is not  convective, then
the pebbles will fragment inside of the envelope and slowly settle towards the
protoplanet's surface while they regrow to macroscopic sizes, while (ii) if the gas envelope is convective, then the
large-scale convection flows transport the pebbles quickly to
the base of the convection flow -- the pebbles then maintain their size relatively well for protoplanets growing within 10 AU, while low-mass protoplanets accreting late further out can have their pebbles reduced to 10 microns or less. 
More complete studies, including the resolved flow of the gas and a more advanced approach to the radiative energy transfer, fragmentation, coagulation and sublimation equations will nevertheless be needed to fully understand the fate of pebbles accreted by a protoplanet and their detachment at the base of the convection flow.

\acknowledgments

We are grateful to the anonymous referee for a thorough referee report that
helped improving the paper. A.J.\ acknowledges funding from the European
Research Foundation (ERC Consolidator Grant 724687-PLANETESYS), the Knut and
Alice Wallenberg Foundation (Wallenberg Academy Fellow Grant 2017.0287) and the
Swedish Research Council (Project Grant 2018-04867)

\appendix

\section{Higher and lower gas accretion rates}

\begin{figure*}
  \begin{center}
       $\dot{M}_\star=3 \times 10^{-7}\,M_\odot\,{\rm yr}^{-1}$, $\tau=3 \times 10^5\,{\rm yr}$ \\
       \includegraphics[width=0.8\linewidth]{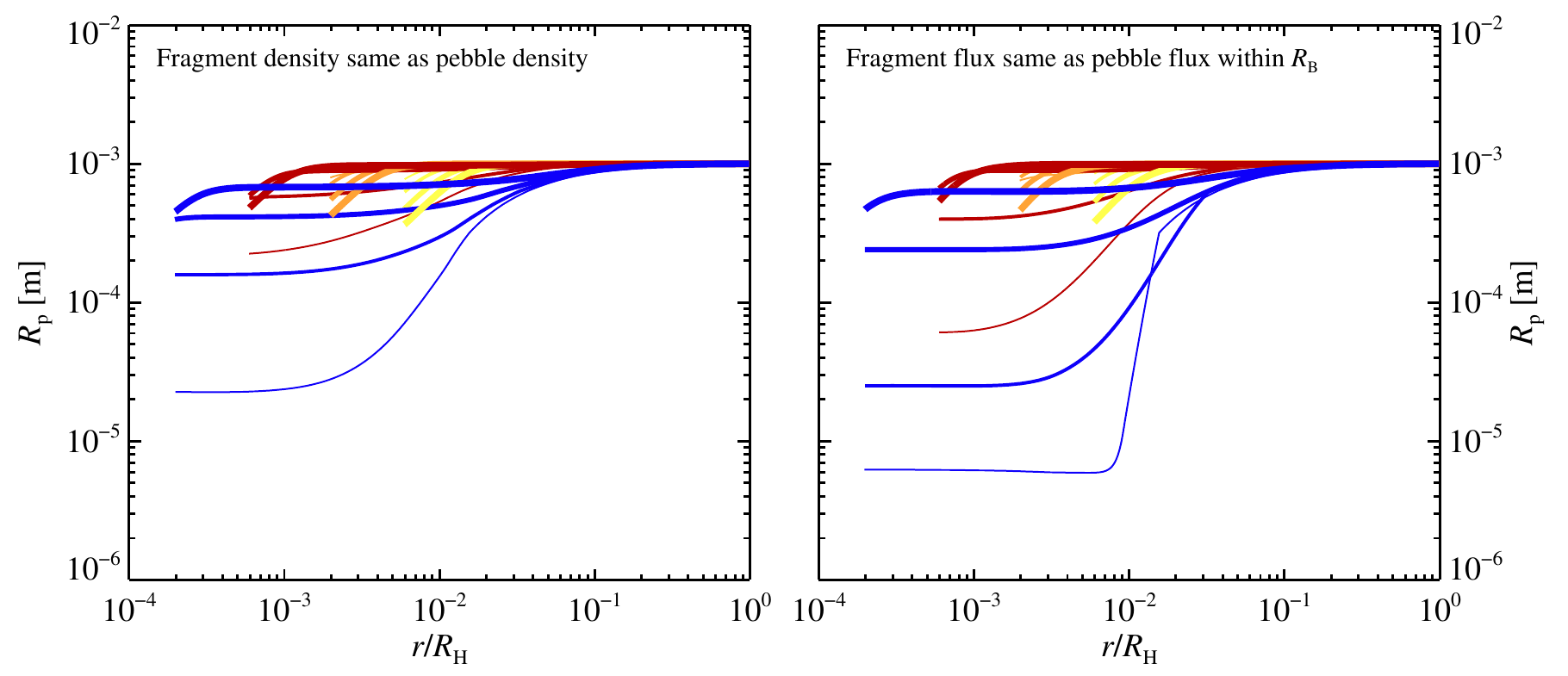}
       \\
       $\dot{M}_\star=3 \times 10^{-7}\,M_\odot\,{\rm yr}^{-1}$, $\tau=3 \times
       10^5\,{\rm yr}$, fully convective \\
       \includegraphics[width=0.8\linewidth]{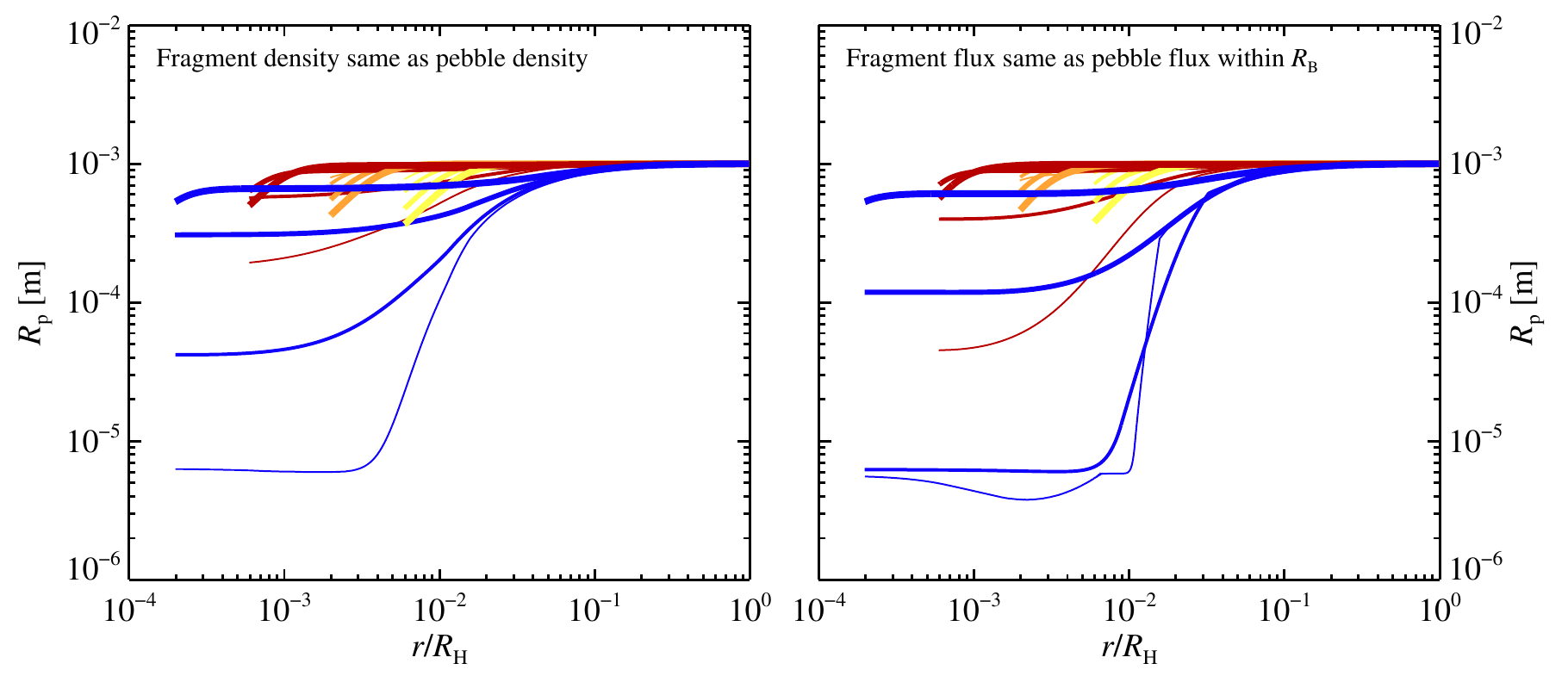}
       \\
       $\dot{M}_\star=1 \times 10^{-8}\,M_\odot\,{\rm yr}^{-1}$, $\tau=1 \times 10^6\,{\rm yr}$ \\
     \includegraphics[width=0.8\linewidth]{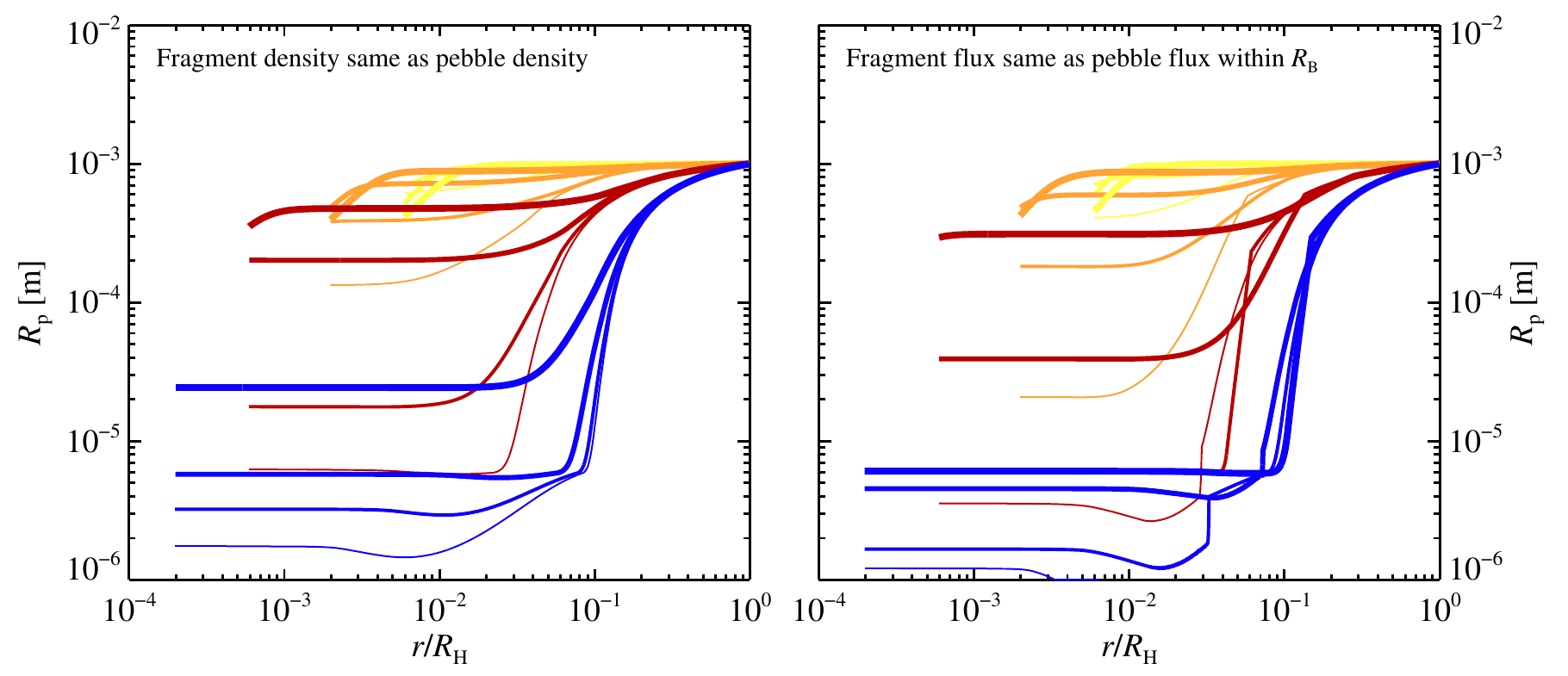}
  \end{center}
  \caption{The pebble size in the models including convective collision speeds and large-scale transport for a
  gas accretion rate of $\dot{M}_\star = 3 \times 10^{-7}\,M_\odot\,{\rm
  yr^{-1}}$ and a pebble accretion time-scale of $\tau=3 \times 10^5\,{\rm yr}$
  (top), for a fully convective model with constant, high opacity (middle) and
  for a lower gas accretion rate $\dot{M}_\star = 10^{-8}\,M_\odot\,{\rm
  yr^{-1}}$ and the nominal pebble accretion time-scale of $\tau=10^6\,{\rm yr}$
  (bottom). The higher gas density leads to slower sedimentation and collision
  speeds and better pebble survival. The fully convective model gives a
  very similar result, although the slightly lower gas density leads to lower
  pebble sizes for the low-mass protoplanets far from the star. The model
  with lower gas accretion rate shown in the bottom panel has a much stronger
  effect on the pebble size than considering fully convective envelopes.}
  \label{f:pebbles_size_sedimentation_convection_Mdot}
\end{figure*}

In this Appendix we demonstrate how the evolutionary stage of the protoplanetary
disc affects the survival of the pebbles in the convective models. We consider
first an earlier evolution stage where the gas accretion rate through the
protoplanetary disc is $\dot{M}_\star=3 \times 10^{-7}\,M_\odot\,{\rm yr}^{-1}$,
three times larger than in the nominal model, with corresponding pebble
accretion time-scale $\tau=3 \times 10^5\,{\rm yr}$ three times smaller than in
the nominal model. The resulting pebble sizes are shown in the top panel of
Figure \ref{f:pebbles_size_sedimentation_convection_Mdot}. The pebbles retain
larger sizes on the way down to the protoplanet compared to the nominal model
(Figure \ref{f:size_sedimentation_convection}), due to the higher gas density
and hence lower sedimentation and collision speeds. We also consider here a
variation of the high-accretion model where the opacity is constant and high
enough that the envelope becomes fully convective. The middle panel of Figure
\ref{f:pebbles_size_sedimentation_convection_Mdot} shows that this has only
little effect on the pebble sizes, since the high-accretion rate case is already
close to fully convective.

Our second parameter variation considers a later growth stage where the gas
accretion rate through the protoplanetary disc is $\dot{M}_\star=1 \times
10^{-8}\,M_\odot\,{\rm yr}^{-1}$, three times lower than in the nominal model.
The lower gas densities lead to an increased destruction of the pebbles.

\section{Constant fragment density}

\begin{figure*}
  \begin{center}
    \includegraphics[width=0.9\linewidth]{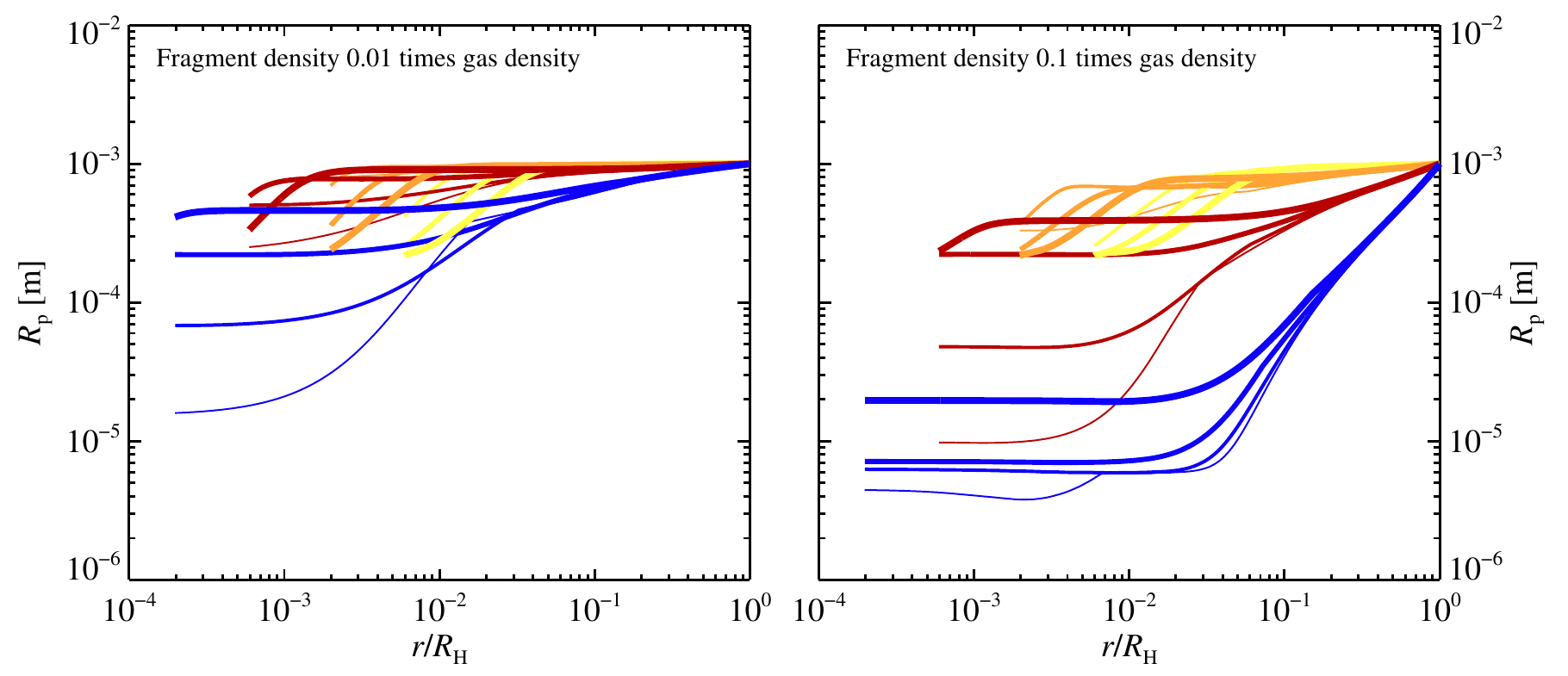}
  \end{center}
  \caption{The pebble size in convective models when fixing the fragment density to either 0.01 (left plot) or 0.1 (right plot) times the gas density. The latter may represent the value needed for the envelope to be fully convective, according to \cite{Ali-DibThompson2019}. The presence of 10\% fragments in the gas leads to lower pebble sizes, with a characteristic size above 100 $\mu$m within 1 AU and falling down to 5-50 $\mu$m further from the star.}
  \label{f:pebbles_size_sedimentation_convection_epsfrag}
\end{figure*}

The density and temperature profiles of the envelope were constructed in the main paper under the assumption of the opacity being provided by a constant dust-to-gas ratio of $0.01$ and using the \cite{BellLin1994} opacity power laws. For the considered pebble accretion rates, the envelopes are close to fully convective (Figure \ref{f:envelope}). We nevertheless did not consider a fully self-consistent coupling between the pebbles, their fragments and the dust opacity. \cite{Ali-DibThompson2019} derived a criterion for the dust mass loading needed to make the envelope fully convective (their equation 33). Here we perform an additional numerical experiment where we fix the fragment density to 0.01 or 0.1 times the gas density. We present the results in Figure \ref{f:pebbles_size_sedimentation_convection_epsfrag}. Increasing the fragment density leads to a decrease in the characteristic pebble size, but the pebbles still maintain sizes above 100 $\mu$m within 1 AU and 5-50 $\mu$m further out.
\end{document}